\documentclass[copyright,creativecommons]{eptcs}
\providecommand{\event}{QAPL 2013}

\usepackage[utf8]{inputenc}
\usepackage[T1]{fontenc}
\usepackage{amsmath,amssymb,stmaryrd,txfonts}
\usepackage{graphicx}
\usepackage{url}
\usepackage{array}
\usepackage{hyperref}
\usepackage[all]{xy}
\usepackage{tikz}
\usepackage{wrapfig,subfigure}
\usepackage{gastex}
\usepackage[strings]{underscore} 
\usepackage{undertilde}
\usepackage{breakurl}    

\newtheorem{defi}{Definition}
\newtheorem{remark}{Remark}
\newtheorem{theo}{Theorem}
\newtheorem{prop}{Proposition}
\newtheorem{lemm}{Lemma}
\newtheorem{coro}{Corollary}
\newtheorem{exam}{Example}

\newenvironment{definition}{\begin{defi} \rm }{\end{defi}}
\newenvironment{theorem}{\begin{theo} \rm }{\end{theo}}
\newenvironment{proposition}{\begin{prop} \rm }{\end{prop}}
\newenvironment{lemma}{\begin{lemm} \rm }{\end{lemm}}

\newenvironment{example}{\begin{exam} \rm }{\end{exam}}

\title{Modal Specifications for Probabilistic~Timed~Systems}
\author{Tingting Han$^1$\quad Christian Krause$^2$
\institute{1. Department of Computer Science, \\University of Oxford }
\email{\{firstname.lastname\}@cs.ox.ac.uk}
\and
Marta Kwiatkowska$^1$\quad Holger Giese$^2$
\institute{2. Hasso Plattner Institute, \\University of Potsdam}
\email{\quad \{firstname.lastname\}@hpi.uni-potsdam.de}
}
\def\titlerunning{Modal Specifications for Probabilistic~Timed~Systems}
\def\authorrunning{T. Han, C. Krause, M.Z. Kwiatkowska \& H. Giese}

\begin{document}

\newcommand{\blankline}{\vspace{\baselineskip}}
\newcommand{\halflineup}{\vspace{-0.5\baselineskip}}

\newcommand{\todo}[1]{\emph{\textcolor{red}{\textbf{TODO:} #1}}}
\newcommand{\note}[1]{\marginpar{\vspace{0.5cm}\scriptsize \textcolor{blue}{\textbf{#1}}}}
\newcommand{\comment}[1]{}

\newcommand{\la}{\ensuremath{\langle}}
\newcommand{\ra}{\ensuremath{\rangle}}
\newcommand{\lb}{\ensuremath{\llbracket}}
\newcommand{\rb}{\ensuremath{\rrbracket}}
\newcommand{\R}{\ensuremath{\mathbb{R}}}
\newcommand{\B}{\ensuremath{\mathbb{B}}}
\newcommand{\N}{\ensuremath{\mathbb{N}}}
\newcommand{\A}{\ensuremath{\mathsf{A}}}
\newcommand{\C}{\ensuremath{\mathsf{C}}}
\newcommand{\Dist}{\ensuremath{\mathit{Dist}}}
\newcommand{\AP}{\ensuremath{\mathit{AP}}}
\newcommand{\Eval}{\ensuremath{\mathit{Eval}}}
\newcommand{\Sat}{\ensuremath{\mathit{Sat}}}
\newcommand{\musttrans}[1]{\ensuremath{\xrightarrow{#1}}}
\newcommand{\maytrans}[1]{\ensuremath{\overset{#1}{\dashrightarrow}}}
\newcommand{\trans}[1]{\ensuremath{\overset{#1}{\rightsquigarrow}}}
\newcommand{\mL}{\ensuremath{\mathcal{L}}}
\newcommand{\mS}{\ensuremath{\mathcal{S}}}
\newcommand{\mX}{\ensuremath{\mathcal{X}}}
\newcommand{\mT}{\ensuremath{\mathcal{T}}}
\newcommand{\mP}{\ensuremath{\mathcal{P}}}
\newcommand{\mM}{\ensuremath{\mathcal{M}}}
\newcommand{\mN}{\ensuremath{\mathcal{N}}}
\newcommand{\mA}{\ensuremath{\mathcal{A}}}
\newcommand{\mR}{\ensuremath{\mathcal{R}}}
\newcommand{\mD}{\ensuremath{\mathcal{D}}}
\newcommand{\mK}{\ensuremath{\mathcal{K}}}
\newcommand{\mG}{\ensuremath{\mathcal{G}}}
\newcommand{\mC}{\ensuremath{\mathcal{C}}}
\newcommand{\mE}{\ensuremath{\mathcal{E}}}
\newcommand{\sA}{\ensuremath{\mathsf{A}}}
\newcommand{\sM}{\ensuremath{\mathsf{M}}}
\newcommand{\sR}{\ensuremath{\mathsf{R}}}
\newcommand{\fR}{\ensuremath{\mathfrak{R}}}
\newcommand{\sT}{\ensuremath{\mathsf{T}}}
\newcommand{\inv}{\mathit{inv}}
\newcommand{\prob}{\mathit{prob}}
\newcommand{\TSteps}{\mathit{TSteps}}
\newcommand{\Steps}{\mathit{Steps}}
\newcommand{\edge}{\mathit{edge}}
\newcommand{\edges}{\mathit{edges}}
\newcommand{\enab}{\mathit{enab}}
\newcommand{\scf}{\mathit{scf}}
\newcommand{\last}{\mathit{last}}
\newcommand{\Path}{\mathit{Path}}
\renewcommand{\inf}{\mathit{inf}}
\newcommand{\Prob}{\mathit{Prob}}
\newcommand{\Succ}{\mathit{Succ}}
\newcommand{\tick}{\mathit{tick}}
\newcommand{\bl}{\medbullet}
\newcommand{\wh}{\medcirc}
\newcommand{\always}{\square}
\newcommand{\eventually}{\Diamond}
\newcommand{\Timediv}{\mathsf{Timediv}}
\newcommand{\mayset}{\mathit{ms}}
\newcommand{\Sd}{\textsf{Sd}}
\newcommand{\Pd}{\textsf{Pd}}
\newcommand{\true}{\mathit{true}}
\newcommand{\false}{\mathit{false}}
\newcommand*{\ltlmodels}{%
\mathrel{\vcenter{\offinterlineskip
\hbox{\scalebox{0.4}{\phantom{a}LTL}}\vskip-.35ex\hbox{$\models$}}}}

\newcommand{\betaA}{{\beta(\mA)}}
\renewcommand{\iff}{\Longleftrightarrow}
\newcommand{\quo}{\varoslash}

\newcommand{\product}{\parallel}

%

\maketitle
\vspace{-0.3cm}
\begin{abstract}

Modal automata are a classic formal model for com\-ponent-based systems that comes equipped with a rich specification theory supporting abstraction, refinement and compositional reasoning. In recent years, quantitative
variants of modal automata were introduced for specifying and reasoning about component-based designs for embedded and mobile systems.
These respectively generalize modal specification theories for timed and
probabilistic systems. In this paper, we define a modal specification language for combined probabilistic timed systems, called \emph{abstract probabilistic timed automata}, which generalizes
existing formalisms. We introduce appropriate syntactic and semantic refinement notions and discuss consistency of our specification language, also with respect to time-divergence. We identify a subclass
of our models for which we define the fundamental operations for abstraction,
conjunction and parallel composition, and show several compositionality results. 

\end{abstract}

\section{Introduction}\label{sec:intro}


The design of complex embedded systems can be supported by component-based design methodologies, which can take the form of specification theories that provide the notions of abstraction and refinement, as well as a rich collection of compositional operators.
A classical and widely used specification theory for component-based design is that of
modal automata~\cite{LarsenT88}\nocite{LarsenNW07}.
A modal automaton is essentially a deterministic automaton equipped with \emph{may}- and \emph{must}-transitions,
which are respectively used to specify allowed and required behavior.
Several technical aspects of modal automata have been studied in the literature, including modal vs.\ thorough refinement,
consistency, abstraction, as well as operators for parallel composition of components and conjunction, where the latter supports independent development. These notions
enjoy a number of important properties, e.g., that conjunction is the greatest lower bound w.r.t.\ modal refinement
and that abstraction is compositional.
In this way, modal automata provide mathematical foundations for designing
and reasoning about component-based systems at the abstract level of interfaces and to derive properties on the implementation level of the global
system. 

%

In recent years, much attention has been dedicated to formulating quantitative extensions of modal automata, for example to support the development of component-based systems which feature real-time and/or probabilistic behavior. An example of these developments are a modal specification language for timed systems called \emph{Modal Event-Clock Specifications} (MECS)~\cite{Bertrand2012}. MECS are essentially
a modal extension of \emph{Event-Clock Automata} (ECAs)~\cite{AFH99}, which form a strict subclass
of the classical \emph{Timed Automata}~\cite{Alur94} model.
Restricting to this model allows Bertrand et al.\ in~\cite{Bertrand2012} to lift a 
number of compositionality properties known for modal automata to the timed setting, i.e., to the model of MECS.
%
%
Another recent quantitative variant are \emph{Abstract Probabilistic Automata}~\cite{DKLLPSW11} (APAs).
While MECS are used to specify timed behavior, APAs enable the specification of abstract probabilistic behavior
using probability constraints. Probabilistic behavior is commonly required for quantifying the likelihood of events,
such as message loss in unreliable channels, or is exploited in the design of randomized protocols. Similarly to MECS,
APAs are equipped with notions for conjunction and parallel composition, as well as a number of compositionality results
that can be used for compositional reasoning and abstraction. However, in many settings, such as in embedded and mobile
systems, a combination of probabilistic and timed behavior is required, which is not supported by APAs and MECS.
Although the specification of combined probabilistic and timed behavior is possible with \emph{Probabilistic Timed Automata}~\cite{KNSS02} (PTAs),
there are no corresponding notions of modalities and abstract probabilistic behavior for PTAs.

In this paper, we introduce a modal specification language for probabilistic timed systems, called
\emph{Abstract Probabilistic Timed Automata} (APTAs), and a subclass of them, called
\emph{Abstract Probabilistic Event-Clock Automata} (APECAs).
APTAs serve as a modal specification language for systems with nondeterministic, probabilistic and timed behavior and support the abstract definition of underspecified probabilistic behavior using constraints (as in APAs). Modalities in the form of may- and must-edges are used to distinguish between allowed and required behavior. APTAs are regarded as specifications which are \emph{implemented} by PTAs. In terms of expressiveness, APTAs subsume PTAs, MECS, and APAs.
Applications of APTAs can be found in the area of component-based systems with real-time and probabilistic behavior, 
e.g., in communication and network protocols for embedded and multimedia systems. 
As a specific example, Stoelinga et al.\ considered PTAs for modeling 
the root contention protocol of the IEEE 1394 standard~\cite{SV99}. The authors defined
several intermediate automata in between the implementation and the specification automaton 
which are related by simple refinement notions. This case study could benefit from modeling 
using APTAs that we introduce here because of their support for abstraction, refinement and compositional operations.

We show the following important results for our models.
For APTAs, we define several appropriate refinement notions and establish a hierarchy among them.  For deterministic APTAs, we show that three of these refinement notions coincide. We provide a consistency check for APTAs based on a reduction to stochastic two-player games. Both probabilistic and strict time-divergence are considered in consistency and refinement checking. We introduce APECAs as a subclass of APTAs and develop abstraction techniques and a compositional theory for this model. In particular, we show that an APECA is related with its abstractions by means of modal refinements. We define conjunction and parallel composition for APECAs, and show that they interact well with modal refinement and abstraction.
Specifically, we show that conjunction is the greatest lower bound after pruning, and that modal refinement is a precongruence with respect to parallel composition. We further show that component-wise abstraction is as powerful as applying the combination of the local abstractions to the entire model.
To the best of our knowledge, this is the first compositional modal specification and abstraction theory for probabilistic timed systems.
Besides the integration of abstract probabilistic behavior, the work in this paper extends~\cite{Bertrand2012} by including a consistency check and refinement relations that consider time divergence.
Moreover, our notion for abstraction non-trivially extends the corresponding APA concept in~\cite{DKLLPSW11} by taking into account the guards of transitions.

\halflineup
\paragraph{Related work.}
This paper is part of an effort to develop a compositional specification theory and assume-guarantee reasoning for component-based systems. 
Previously, we have developed a linear-time specification theory for components~\cite{ChenCJK12} and its timed extension~\cite{ChiltonKW12}. 
We have also formulated the corresponding sound and complete compositional assume-guarantee rules~\cite{CJK12}, demonstrating their application on examples of component-based systems from the networking domain.
Linear-time refinement for probabilistic systems is known not to be compositional, and hence we focus on modal specifications.
Modal specification theories for probabilistic systems include
APAs~\cite{DKLLPSW11} and Constraint Markov Chains~\cite{Caillaud2011}. A specification theory for real-time systems is defined in~\cite{David2010} including a set of operators supporting stepwise design of timed systems.
A general approach for quantitative specification theories with modalities is presented in~\cite{Bauer12}.
A robust specification theory for Modal Event-Clock Automata is discussed in~\cite{FL12}. And aggressive abstraction techniques for probabilistic automata are explored in \cite{SK12}.
\halflineup
\paragraph{Structure.}
In Section~\ref{sec:prelims}, we recall relevant notions for APAs and PTAs.
Section~\ref{sec:apta} introduces our new model of Abstract Probabilistic Timed Automata.
In Section~\ref{sec:refinement}, refinement notions for APTAs are defined and compared.
Section~\ref{sec:abst} is devoted to abstraction for APTAs.
In Section~\ref{sec:comp}, we define conjunction and parallel composition, and present compositionality results for APECAs (a strict subclass of APTA).
Section~\ref{sec:conclusions} concludes and discusses future work.

\section{Preliminaries}
\label{sec:prelims}

In this section, we recall important definitions for PTAs~\cite{KNSS02} and refer the reader to \cite{DKLLPSW11} and \cite{HKKG13} 
for APAs. 
%
We first recall some elementary notions. A \emph{discrete probability distribution} over a denumerable 
set $S$ is a function $\mu : S \to [0,1]$ with $\sum_{s\in S} \mu(s) = 1$. 
The set of all discrete probability distributions over $S$ is denoted 
by $\Dist(S)$. For a given $s\in S$, the \emph{point distribution} $\mu_s$ is the unique distribution 
on $S$ with $\mu_s(s)=1$. 
We denote by $\R_+$ the set of non-negative reals. Let $\B_2=\{\bot,\top\}$ 
and $\B_3 = \{ \bot, ?, \top\}$ be the complete lattices with the respective orderings 
$\bot < \top$ and $\bot < \; ? < \top$, and meet ($\sqcap$) and join ($\sqcup$) operators.

\subsection{Probabilistic Timed Automata}
\label{sec:pta}

We now recall the standard timed automata notions of clock valuations and guards.
For a finite set $X$ of \emph{clocks}, a \emph{clock valuation} is a function $v: X \to \R_+$. The set of all clock valuations over $X$ is denoted by $\R_+^X$. For any $v \in\R_+^X$ and $t\in\R_+$, we use $v+t$ to denote the clock valuation defined as $(v+t)(x)=v(x)+t$ for all $x\in X$. We use $v[Y:=0]$ to denote the clock valuation obtained from $v$ by resetting all of the clocks in $Y\subseteq X$ to 0, and leaving the values of all other clocks unchanged; formally, $v[Y:=0](x)=0$ if $x\in Y$ and $v[Y:=0](x)=v(x)$ otherwise. We write $\overline{0}$ for the clock valuation that assign 0 to all clocks.

Let $X = \{x_1,\ldots,x_n\}$ be a set of clocks. A \emph{clock constraint} or \emph{guard} $g$ on $X$ is an expression of the form $x \sim c$ 
such that $x,y\in X$, $c \in \R_+$ and $\sim \;\in \{ \leq, <, >, \geq \}$, or a conjunction of guards. A clock valuation~$v$ satisfies~$g$, written as $v\triangleright g$, iff $g$ evaluates to true when all clocks $x \in X$ are substituted with their clock value $v(x)$. Let $CC(X)$ denote the set of all guards over $X$, and let $CC_N(X)$ denote the set of guards on $X$ involving expressions with constants less or equal to $N$, where $N$ is the maximal constant in all guards.

\begin{definition}[PTA~\cite{KNSS02}] A \emph{probabilistic timed automaton} is a 
tuple $\mM=(L,A,X,AP,V,T,l_0)$ where $L$ is a finite set of locations with initial location $l_0\in L$; 
$A$ is a finite set of actions; $X$ is a finite set of clocks; $AP$ is a finite set of atomic propositions; 
$V:L\to 2^{\AP}$ assigns atomic propositions to locations; and 
$T:L\times CC(X)\times A\times\Dist(2^X\times L)\to \B_2$ is a probabilistic edge function.
\end{definition}
We write $l\musttrans{g,a}\mu$ iff $T(l,g,a,\mu)=\top$, which comprises a source location $l$, a guard $g$, and a probability distribution $\mu$ which assigns probabilities to pairs of the form $(Y,l')$, where $Y\subseteq X$ is a set of clocks to be reset and $l'$ is a target location. The behavior of a PTA is as follows: in any location 
a probabilistic edge can be taken if its guard is satisfied by the current values of the clocks. Once a probabilistic edge is nondeterministically selected, the choice for a particular target location and set of clocks to be reset is made probabilistically using $\mu$.

We define the semantics of a PTA $\mM$ by mapping it to a probabilistic automaton $\sM$ by employing the classical region equivalence for timed automata~\cite{Alur94}. 
A \emph{probabilistic automaton} (PA)~\cite{Seg95,DKLLPSW11} $\sM = (S,A,AP,V,T,s_0)$ consists of a set of states $S$ with initial state $s_0$, 
a set of actions $A$, a set of atomic propositions $AP$, a valuation function $V: S \to 2^{AP}$ and 
a probabilistic transition function $T: S \times A \times \Dist(S) \to \B_2$.
A \emph{region} $\theta$ is the set of clock valuations which satisfy exactly the same guards of $CC_N(X)$. Given a region $\theta$, we write $\Succ(\theta)$ for the union of all regions that can be obtained from $\theta$ by letting time elapse. Given a guard $g\in CC(X)$, we write $\theta\subseteq g$ iff for all valuations $v\in \theta$ it holds that $v\triangleright g$. We denote the set of all regions by $\Theta_N(X)$ and simply write $\Theta$ if clear from context. 
We now define the semantics of a PTA in terms of a PA. 

\begin{definition}[Region PA]\label{def:region-automaton}
For a given PTA $\mM=(L,A,X,AP,V,T,l_0)$, the associated \emph{region PA} is given by $\sR(\mM) = (S, A', AP, V', T', s_0)$ where $S = L \times \Theta$ and $s_0 = (l_0,\overline{0})$, $A' = \Theta \times A \times (2^X)^S$, $V'(l,\theta) = V(l)$, and $T'$ is induced by $T$ in the following way: for any $l\in L$ and any $\theta\in \Theta$ such that $(l,\theta)$ is reachable from $(l_0,\overline{0})$, if $l\musttrans{g,a}_{\mM}\mu$ then for each region $\theta'' \in \Succ(\theta) \cap g$ there exists $\zeta: S \to 2^X$ and $\mu'\in \Dist(S)$ such that $(l,\theta)\musttrans{\theta'',a,\zeta}_{\sR(\mM)}\mu'$ and:\\ 
\halflineup
$$\zeta(l',\theta') = Y \mbox{ and }\mu'(l',\theta') = \mu(l',Y) \mbox{ if }\theta' = \theta''[Y:=0]; \mbox{ and }
\zeta(l',\theta') = \emptyset \mbox{ and }\mu'(l',\theta') = 0 \mbox{ o.w.}$$
\end{definition}
In the derived region PA, the transition labels $(\theta'',a,\zeta) \in A'$ consist of the region $\theta''$ that represents the time window in which the transition is taken, the fired action $a \in A$, and for each target state $s \in S$ the set of clocks $\zeta(s) \subseteq X$ that are being reset when $s$ is probabilistically chosen. 
W.l.o.g., we assume that $\sR(\mM)$ is always pruned, i.e., all its states are reachable. 

Similarly as shown for timed specifications in~\cite{Bertrand2012}, any PA that is defined over the alphabet $\Theta\times A \times (2^X)^S$ can be interpreted as a PTA again. Intuitively, the states of the PA are interpreted as the locations of the corresponding PTA and the information about the guards, actions and clock resets for edges is derived from the transition labels of the PA. We introduce the operator $\mT$ which translates any given PA $\sM$ over the alphabet $\Theta\times A \times (2^X)^S$ into the PTA $\mT(\sM)$. The application of $\mT \circ \sR$ allows us, moreover, to define a normal form for PTAs.                                                                                                                                                                                                                                                                                                                                                                                                                                                                                                                                                                                                                                                                                                                                                                                                                                                                                                                                                                                                                                                                                                                                                                                                                                                                                                                                                                                                                                                                                                                                                                                                                                                                                                                                                                                                                                                                                                                                                                                                                                                                                                                                                                                                                                                                                                                                                                                                                                                                                                                                                                                                                                                                                                                                                                                                                                                                                                                                                                                                                                                                                                                                                                                                                                                                                                                                                                                                                                                                                                                                                                                                                                                                                                                                                                                                                                                                                                                                                                                                                                                                                                                                                                                                                                                                                                                                                                                                                                                                                                                                                   
\begin{definition}[Normal form]
A PTA $\mM$ is in \emph{normal form} iff it is isomorphic to (the reachable part) of $(\mT {\circ} \sR)(\mM)$.
\end{definition}
Note that, if a PTA in normal form, every location is associated with a unique region. Moreover, $(\mT{\circ} \sR)(\mM)$ is isomorphic to $(\mT{\circ} \sR)^2(\mM)$ for any PTA $\mM$. The PTA in normal form in needed later for technical reasons (e.g., Proposition\,\ref{pro:pa-apta} or Theorem\,\ref{thm:apta-apa-reduction}).

\section{Abstract Probabilistic Timed Automata}
\label{sec:apta}

We now define \emph{Abstract Probabilistic Timed Automata} (APTAs) as the central model of this paper. APTAs extend PTAs in three ways: (1) probability distributions are generalized to probability constraints, (2) may- and must-transitions are distinguished, and (3) locations are labeled with sets of admissible atomic propositions. All three modeling concepts are borrowed from APAs~\cite{DKLLPSW11}. We use satisfaction relations to relate APTAs with PTAs that implement them. Let a \emph{probability constraint} $\varphi$ be a symbolic representation of a set of probability distributions 
over a set $S$. As in~\cite{DKLLPSW11}, we do not fix the language for probability constraints. 
The set of probability distributions that satisfy $\varphi$ is denoted by $\Sat(\varphi) \subseteq \Dist(S)$. 
We define the constraints $\true$ and $\false$, for which we require $\Sat(\true) = \Dist(S)$ and 
$\Sat(\false) \!=\! \emptyset$. The set of probability constraints over $S$ is denoted by $PC(S)$.

\begin{definition}[APTA]
An \emph{abstract probabilistic timed automaton} is a tuple $\mA = (L,A,X,$ $AP,V,T,l_0)$, where $L$, $A$, $X$, $AP$, $l_0$ are defined as for PTAs; $V: L \to 2^{2^{AP}}$ assigns sets of admissible atomic propositions to locations;
and $T: L {\times} CC(X) {\times} A {\times} PC(2^{X} {\times} L) {\to} \B_3$ is a three-valued probabilistic edge function.
\end{definition}
\halflineup
We use the notation $l\maytrans{g,a}\varphi$ to denote may-edges (formally if $T(l,g,a,\varphi)=\;?$), $l\musttrans{g,a}\varphi$ for must-edges (if $T(l,g,a,\varphi)=\top$), 
and $l\trans{g,a}\varphi$ for may- or must-edges, where $g\in CC(X)$ is a guard, $a\in A$ is an action, and $\varphi \in PC(2^{X} \times L)$ is a 
probability constraint. 

\begin{example}
Fig.\,\ref{fig:scheduler} depicts an example APTA modeling a scheduler component. In the initial location $l_0$, tasks can be submitted to the scheduler and will be started within $1$ time unit. Two types of tasks can occur: short- and long-running ones. The choice between them is probabilistic according to the probability constraint $\varphi_p= (0.25\le p_1\le 0.75)\,\wedge (0.25\le p_2 \le 0.75)\,\wedge (p_1+p_2=1)$. Tasks either finish in the expected time frame or can be canceled at any point. The canceling of tasks is modeled using may-edges, and thus is not required to be realized by implementations.
\end{example}
\begin{figure}[h]
\vspace{-2mm}
\scalebox{0.5}{
\begin{large}\begin{picture}(136,60)(-15,-64)\nullfont
\node[NLangle=0.0](n26)(16.0,-40.0){$l_1$}

\nodelabel[ExtNL=y,NLangle=-90.0,NLdist=0.9](n26){$\{\{\mathit{idle}\},\{\mathit{busy}\}\}$}

\node[NLangle=0.0](n27)(72.0,-26.0){$l_2$}

\nodelabel[ExtNL=y,NLangle=-75.0,NLdist=0.9](n27){$\{\{\mathit{short}\}\}$}

\node[NLangle=0.0](n29)(72.0,-54.0){$l_3$}
\nodelabel[ExtNL=y,NLangle=-90.0,NLdist=0.9](n29){$\{\{\mathit{long}\}\}$}

\node[NLangle=0.0,iangle=0,Nmarks=i](n30)(124.0,-40.0){$l_0$}
\nodelabel[ExtNL=y,NLangle=-40,NLdist=0.9](n30){$\{\{\mathit{idle}\}\}$}

\node[Nfill=y,fillcolor=Black,Nw=2.0,Nh=2.0,Nmr=1.0](n31)(51.88,-40.0){}

\node[Nw=0.0,Nh=0.0,Nmr=0.0](n32)(124.0,-10.0){}

\node[Nw=0.0,Nh=0.0,Nmr=0.0](n33)(16.0,-10.0){}

\drawedge[AHnb=0](n26,n31){ $\;\;\mathit{start},\, $ $0\le x<1$}

\drawedge(n31,n27){$p_1$}

\drawedge[ELside=r,ELdist=1.77](n31,n29){$p_2$}

\drawedge[AHnb=0](n32,n30){ }

\drawedge[ELside=r,ELdist=2.12,AHnb=0](n32,n33){$\mathit{submit},\, $ $ x:=0$}

\drawedge(n33,n26){}

\drawedge[ELdist=2.18,curvedepth=6,ELdist=0.8](n27,n30){$\mathit{finish},\, $ $ 0<x\le 2$}

\drawedge[ELside=r,dash={2.0 2.0 2.0 3.0}{0.0},curvedepth=3](n27,n30){\emph{cancel}}

\drawedge[ELside=r,ELdist=0.4,curvedepth=-6](n29,n30){$\mathit{finish},\, $ $  2<x\le 10$}

\drawedge[dash={2.0 2.0 2.0 3.0}{0.0},ELdist=1.38,curvedepth=-3](n29,n30){\emph{cancel}}


\end{picture}\end{large}
}\qquad\quad\scalebox{0.5}{
\begin{large}\begin{picture}(191,60)(-10,-64)\nullfont
\node[NLangle=0.0](n26)(16.0,-40.0){$l_1$}

\nodelabel[ExtNL=y,NLangle=-90.0,NLdist=0.9](n26){$\{\mathit{busy}\}$}

\node[NLangle=0.0](n27)(72.0,-18.0){$l_2$}

\nodelabel[ExtNL=y,NLangle=-50.0,NLdist=0.9](n27){$\{\mathit{short}\}$}

\node[NLangle=0.0](n28)(72.0,-40.0){$l_3$}
\nodelabel[ExtNL=y,NLangle=-90.0,NLdist=0.9](n28){$\{\mathit{long}\}$}

\node[NLangle=0.0](n29)(72.0,-62.0){$l'_3$}
\nodelabel[ExtNL=y,NLangle=-90.0,NLdist=0.9](n29){$\{\mathit{long}\}$}

\node[NLangle=0.0,iangle=0,Nmarks=i](n30)(124.0,-40.0){$l_0$}
\nodelabel[ExtNL=y,NLangle=-40,NLdist=0.9](n30){$\{\mathit{idle}\}$}

\node[Nfill=y,fillcolor=Black,Nw=2.0,Nh=2.0,Nmr=1.0](n31)(51.88,-40.0){}

\node[Nw=0.0,Nh=0.0,Nmr=0.0](n32)(124.0,-8.0){}

\node[Nw=0.0,Nh=0.0,Nmr=0.0](n33)(16.0,-8.0){}

\drawedge[AHnb=0](n26,n31){ $\mathit{start},\, $ $0\le x<1$}

\drawedge(n31,n28){$0.3$}

\drawedge(n31,n27){$0.4$}

\drawedge[ELside=r,ELdist=1.77](n31,n29){$0.3$}

\drawedge[AHnb=0](n32,n30){ }

\drawedge[ELside=r,ELdist=2.12,AHnb=0](n32,n33){$\mathit{submit},\, $ $ x:=0$}

\drawedge(n33,n26){}

\drawedge[ELdist=2.82, ELdist=0.3](n28,n30){$\mathit{finish},\, $ $ 2<x\le 6$}

\drawedge[ELdist=2.18,curvedepth=6,ELdist=0.8](n27,n30){$\mathit{finish},\, $ $ 0<x\le 2$}

\drawedge[ELside=r,ELdist=0.4,curvedepth=-6](n29,n30){$\mathit{finish},\, $ $  6<x\le 10$}

\drawedge[ELdist=1.38,curvedepth=-3](n29,n30){\emph{cancel}}

\end{picture}\end{large}
}\vspace{3mm}\\
\begin{minipage}{0.47\textwidth}
\caption{An example APTA specification for a scheduler component.} \label{fig:scheduler}
\end{minipage}\qquad\begin{minipage}{0.47\textwidth}
 \caption{An example PTA implementing the APTA in Fig.\,\ref{fig:scheduler}.}\label{fig:impl}
\end{minipage}\vspace{-2ex}
\end{figure}

\begin{definition}[APTA Satisfaction] 
\label{def:satisfaction2}
Let $\mM = (L, A, X, AP, V, T, l_0)$ be a PTA in normal form and $\mA = (L',A,X,AP,V',T',l_0')$ be an APTA. $R\subseteq L\times L'$ is called a \emph{satisfaction relation} iff, for all $(l,l')\in R$, these conditions hold:
 \begin{enumerate}
  \item $\forall a\in A$, $\forall \varphi'\in PC(2^X\times L')$, $\forall g\in CC(X)$ and $\forall \theta\in\Theta$: if 
  $l'\!\musttrans{g,a}_{\mA}\varphi'$ 
  and both $(l,\theta)$ and $(l',\theta)$ are reachable in $\mM$ and $\mA$ respectively, then $\exists n\in\N$, $\exists g_1,\ldots,g_n\in CC(X)$ and $\exists \mu_1,\ldots,\mu_n \in\Dist(2^X\times L)$ with: (i) $\Succ(\theta)\cap g \subseteq \Succ(\theta)\cap \bigcup_{i=1}^n g_i$; and 
(ii) $\forall 1\le i\le n$:  
  $l\!\musttrans{g_i,a}_{\mM}\mu_i$
  and $\exists \mu_i'\in\Sat(\varphi')$ s.t.\ $\mu_i\Subset_R \mu_i'$ (see the definition of $\Subset_R$ in \cite{HKKG13});
\item $\forall a\in A$, $\forall \mu\in\Dist(2^X\times L)$, $\forall g\in CC(X)$: if 
$l \musttrans{g,a}_{\mM} \mu$
then $\exists g'\in CC(X)$ and $\exists \varphi' \in PC(2^X \times L')$: 
$l' \trans{g',a}_{\mA} \varphi'$, $g \subseteq g'$
and $\exists \mu'\in\Sat(\varphi')$ with $\mu\Subset_R \mu'$; 
\item  $V(l)\in V'(l')$.
\end{enumerate}
We say that $\mM$ satisfies $\mA$, denoted $\mM\models\mA$, iff there exists a satisfaction relation relating $l_0$ and $l_0'$. If $\mM\models\mA$, $\mM$ is called an \emph{implementation} of $\mA$.
\end{definition}
Condition 1 states that any must-edge in the specification is required to be realized in an implementation (possibly split up into several edges emitting from one location). Condition 2 ensures that any edge in the implementation is allowed by the specification (as a may- or a must-edge). Note that, since $\mM$ is in normal form, the guard in the edge $l \musttrans{g,a}_{\mM} \mu$ is necessarily a region. The set of all implementations of $\mA$ is given by $\lb\mA\rb=\{\mM\!\mid \!\mM\models\mA\}$. 

\begin{example}
Fig.\,\ref{fig:impl} depicts an implementation of the APTA for a scheduler component in Fig.\,\ref{fig:scheduler}. We indicate the satisfaction relation by using equal location indices, e.g., the locations $l_3$ and $l'_3$ in the implementation are in relation with $l_3$ in the specification. The implementation differs from the specification in the following aspects. After a task has been submitted, the scheduler becomes busy, i.e., the set of atomic propositions $\{ busy \}$ is chosen for location $l_1$. Two types of long-running tasks are distinguished in the implementation: ones that finish in the interval $(2,6]$ and ones that finish in the interval $(6,10]$. Only tasks of the latter type can be canceled. The probability constraint $\varphi_p$ is realized by the probability distribution assigning $0.4$ to $l_2$, and $0.3$ to $l_3$ and $l'_3$, respectively. Note, however, that this PTA is not in normal form, because the location $l_0$ can be reached within three different regions: $(0,2]$, $(2,6]$ and $(6,10]$. Thus, by splitting up location $l_0$, the normal form can be obtained and the satisfaction relation is constructed.
\end{example}

\subsection{The Region-Based Interpretation}

We show now that the check for the existence of a satisfaction relation between a PTA and an APTA can be reduced to a check for a satisfaction relation between their corresponding probabilistic region automata.
Analogously to the mapping of a PTA to a PA using the region construction (cf.\ Def.\,\ref{def:region-automaton}), we can transform any APTA $\mA$ into an APA $\sR(\mA)$. In the resulting APA $\sR(\mA)$, the types of the three-valued edge function are inherited from $\mA$. The transition relation $T_{\sR(\mA)}:S\times\Theta\times A\times (2^X)^S\times PC(S)\to\B_3$ is lifted from distributions to constraints by:
\begin{itemize}
  \item for any $l\in L$, $\theta \in \Theta$ such that $(l,\theta)$ is reachable from $(l_0,\overline{0})$: if $l\musttrans{g,a}_\mA \varphi$ then for each $\theta'' \in \Succ(\theta) \cap g$ there exists $\zeta: S \to 2^X$ and $\varphi' \in PC(S)$ such that $(l,\theta)\musttrans{\theta'',a,\zeta}_{\sR(\mA)}\varphi'$, and $\exists \mu \! \in \! \Sat(\varphi) $ iff $ \exists \mu'\! \in \!\Sat(\varphi')$ with:
\halflineup
$$\zeta(l',\theta') = Y \mbox{ and }\mu'(l',\theta') = \mu(l',Y) \mbox{ if }\theta' = \theta''[Y:=0]; \mbox{ and }
\zeta(l',\theta') = \emptyset \mbox{ and }\mu'(l',\theta') = 0 \mbox{ o.w.}$$\\[-4ex]
and analogously for all may-edges.
\end{itemize}
%
\begin{proposition}\label{pro:pa-apta}
Consider an APTA $\mA = (L,A,X,AP,V,T,l_0)$ and a PA $\sM = (S,A',AP,V',T',s_0)$ where $A' = \Theta(X) \times A \times (2^X)^S$. If $\sM \models \sR(\mA)$ then $\mT(\sM)$ is in normal form and $\mT(\sM) \models \mA$.
\end{proposition}This proposition will be used later in Section \ref{sec:consistency}.
\begin{theorem}\label{thm:apta-apa-reduction}
Given a PTA $\mM= (L, A, X, AP, V, T, l_0)$ in normal form and APTA $\mA= (L',A,X,$ $AP,V',T',l_0')$, and let $\sR(\mM)=(S,A_\sR,AP,V_\sR,T_\sR,s_0)$ and $\sR(\mA)=(S',A_\sR,AP,V_\sR',T_\sR',s_0')$ be the respective region automata. Then
$\mM \models \mA\mbox{ if and only if } \sR(\mM) \models \sR(\mA)$.
\end{theorem}
Theorem~\ref{thm:apta-apa-reduction} does not hold for arbitrary PTAs, since the \emph{if}-part only holds for PTAs in normal form. That is to say, there exist a PTA $\mM$ and an APTA $\mA$ such that $\mM\not\models\mA$, while $\sR(\mM)\models\sR(\mA)$. A similar example for (non-probabilistic) timed modal specifications can be found in \cite{Bertrand2012}.

\begin{definition}[Deterministic APTA]\label{def:det}
Given an APTA $\mA = (L,A,X,AP,V,T,l_0)$ and its region automaton $\sR(\mA) = (S, A', AP, V', T', s_0)$. $\mA$ is called:
\begin{itemize}
  \item \emph{action-deterministic}, iff for all reachable states $s$ in $\sR(\mA)$ it
holds: if there exist 
$s \trans{\theta_1,a,\zeta_1}_{\sR(\mA)} \varphi_1$ and
$s \trans{\theta_2,a,\zeta_2}_{\sR(\mA)} \varphi_2$ such that $\varphi_1\neq \varphi_2$,
 then $\theta_1 \cap \theta_2 = \emptyset$;
  \item \emph{AP-deterministic}, iff $s\trans{\theta,a,\zeta}_{\sR(\mA)} \varphi$ implies that 
  for all $\mu',\mu'' \in \Sat(\varphi)$, and $s'\neq s'' \in S$ it holds: $(\mu'(s')>0 \;\wedge\; \mu''(s'')>0) \Longrightarrow V(s') \,\cap\, V(s'') = \emptyset$.
\end{itemize}
$\mA$ is called \emph{deterministic} iff it is action-deterministic and AP-deterministic.
\end{definition}

\noindent
Note that Def.\,\ref{def:det} is inspired by~\cite{DKLLPSW11}.
Action-determinacy can also be enforced on the syntactical level. 
However, such a definition would only be a sufficient, but
not a necessary condition. 
\vspace{-2mm}

\subsection{Consistency}
\label{sec:consistency}

Consistency of a specification refers to the property that there exists at least one model for this specification. In our setting, an APTA~$\mA$ is said to be consistent if it admits at least one implementation, hence formally iff $\lb \mA \rb \neq \emptyset$. For any given APTA~$\mA$, we can decide whether the APA $\sR(\mA)$ is consistent and, if so, derive a PA $\sM$, such that $\sM \models \sR(\mA)$~\cite{DKLLPSW11}. $\mT(\sM)$ is then a PTA with finitely many states, and, by Proposition~\ref{pro:pa-apta}, a model of $\mA$.  

In order to deal with consistency also on the syntactic level, we further 
define a location $l\in L_\mA$ in an APTA $\mA=(L_\mA,A,X,AP,V_\mA,T_\mA,l_\mA^0)$
to be consistent if 
$V_\mA(l)\neq \emptyset$ and for all guards $g\in CC(X)$, actions $a\in A$ and probability constraints 
$\varphi\in PC(2^X\times L_\mA)$ it holds: if $T_\mA(l,g,a,\varphi)=\top$ then $\Sat(\varphi)\neq\emptyset$. 
Note that inconsistency of a location does not imply inconsistency of the whole APTA.
In order to decide whether an APTA is consistent, we follow the usual approach and 
use a \emph{pruning operator} $\beta$ that filters out distributions leading to inconsistent
locations~\cite{DKLLPSW11}. The detailed definition and properties of a pruning operator can be found in \cite{HKKG13}.

The following theorem shows that the application of $\beta$ operator does not change the set of implementation. And it also implies that if $\beta^*(\mA)$ is empty, then $\mA$ is inconsistent. 
\begin{theorem}\label{thm:consistency}
 For any APTA $\mA$, it holds that $\lb\mA\rb=\lb\beta(\mA)\rb=\lb\beta^*(\mA)\rb$.
\end{theorem}

The above definition of consistency, however, places no restrictions on the derived implementations. In particular, the derived PTA could show unrealistic behaviors by preventing time from diverging. Therefore, we also aim at checking consistency in such a way that only divergent implementations are considered. Note that a divergent consistent APTA must be consistent, therefore we assume that the APTAs that we deal with are already consistent. 

We consider the set of \emph{strict} and \emph{probabilistic divergent} (\textsf{Sd} and \textsf{Pd}, for short) implementations of $\mA$ given by $\lb\mA\rb^{\textsf{Sd}}=\{\mM\mid \mM\models\mA\mbox{ and }\mM\mbox{ is strict divergent}\}$ and $\lb\mA\rb^{\textsf{Pd}}=\{\mM\mid \mM\models\mA\mbox{ and }\mM$ is probabilistic divergent$\}$, respectively. The formal definitions of probabilistic and strict divergency can be found in \cite{Sproston09} and \cite{HKKG13}. 
Now we define an APTA to be \textsf{Sd}- or \textsf{Pd}-\emph{consistent} if it admits at least one \textsf{Sd} or \textsf{Pd} implementation, i.e., $\lb\mA\rb^{\textsf{Sd}} \neq \emptyset$ or $\lb\mA\rb^{\textsf{Pd}} \neq \emptyset$. 

Theorem~\ref{thm:game} shows that a time-divergence sensitive consistency check for APTAs can be defined based 
on a reduction to APAs and stochastic two-player games. 
The details of this technique can be found in \cite{HKKG13}. 
This result effectively allows us to check whether an APTA has at least one strict or probabilistic divergent implementation.

\begin{theorem}\label{thm:game}
An APTA $\mA$ is \textsf{Pd} (resp.\ \textsf{Sd}) consistent
if and only if in the game $\mG(\mA)$, the $\bl$-player has a winning strategy for the objective 
$\mathbb{P}_{=1}(\always\eventually\tick)$ (resp.\ objective $\always\eventually\tick$).
\end{theorem}

\subsection{Abstract Probabilistic Event-Clock Automata}
\label{sec:apeca}

We now introduce \emph{Abstract Probabilistic Event-Clock Automata} $($APECAs$)$, which form a strict subclass of APTA, where clock resets are not arbitrary: 
each action $a$ is associated with a clock $x_a$ which is reset exactly when the action $a$ occurs. This kind of clock resets originated from 
\emph{Event-Clock Automata} (ECAs)~\cite{AFH99}: they 
form a strict subclass of TA, but they enjoy nice properties, e.g., they are closed under union and intersection, 
and can be determinized. 
\begin{definition}[APECA]
 A (complete) \emph{abstract probabilistic event-clock automaton} $($APECA$)$ is a tuple $\mE=(L,A,X_A,AP,V,T,l_0)$, where $L$, $A$, $AP$, $l_0$, and $V$ are defined as for APTAs;
\begin{itemize}
  \item $X_A$ is a set of clocks where every $x_a\in X_A$ corresponds to an action $a\in A$;
 \item $T:L\times CC(X_A)\times A\times PC(L)\to \B_3$ is a three-valued probabilistic transition function, s.t.\ 
 for all $l\in L,a\in A$: $\bigvee_i \{ g_i \mid \exists \varphi_i: T(l,g_i,a,\varphi_i) \neq \bot \} = \true$. 
\end{itemize}
\end{definition}

\begin{example}[APECA]
\label{ex:apeca}
Fig.~\ref{fig:example1} depicts two APECAs $Cl$ and $Acc$. $Cl$ models a clients requesting 
access to a given resource. It can either invoke \textit{get} to 
request the resource; or \textit{grant} to access it. The action \textit{extra} is used 
when a privileged access with extended time is needed. We use $!$ and~$?$ to indicate 
whether an action comes from the designed component or from its environment. The clock 
corresponding to the action \textit{get} is $x_{get}$. The client sends 
a second \textit{get}-request at most one time unit after the first request. 
With a probability satisfying constraint $\varphi_1$, 
the client terminates and stops requesting resources (state 2). The client 
can also request extended time at any moment while it is still active. Let the probability 
from state 1 to state 0 be $p_1$ and from state 1 to state~2 be $p_2$ in $Cl$. 
Then $\varphi_1$ could be defined as $0\le p_1\le 1/3$ and $p_1+p_2=1$.

The APECA $Acc$ specifies the behavior of an access controller. If the access to the resource is granted, then it should happen within 2 time units after reception of a \textit{get} request. 
In case of a privileged access with extra time, this duration will be extended to 
at most 4 time units. However, with a certain probability (satisfying $\varphi_2$), 
the access controller will switch back to the default access time of 2 time units. The 
probability constraint $\varphi_2$ can be defined in a similar way as $\varphi_1$.
The use of probability constraints is explained in more detail in 
Examples~\ref{ex:conjunction} and~\ref{ex:parallel}.


\end{example}

\begin{wrapfigure}{r}{3.4in}
\vspace{-4mm}
\hspace{1cm}\scalebox{0.65}{
\begin{picture}(45,83)(30,-83)
\put(0,-83){}
\node(n1)(23.49,-16.25){0}

\node(n2)(63.24,-16.25){1}

\node(n3)(23.49,-32.25){2}

\drawedge[curvedepth=5.0](n1,n2){!get}

\drawedge[ELpos=70,ELdist=-0.1,curvedepth=5.0](n2,n1){?grant,$\varphi_1$}

\drawedge(n2,n3){}

\drawloop[dash={2.0 2.0 2.0 3.0}{0.0}](n1){!extra }

\drawloop[dash={2.0 2.0 2.0 3.0}{0.0},loopangle=-90.0](n2){!extra }

\drawloop[dash={2.0 2.0 2.0 3.0}{0.0}](n2){!get, $x_{get}\le 1$ }

\node(n4)(23.26,-55.01){0'}

\node(n5)(63.26,-55.01){1'}

\drawloop(n4){?get}

\drawloop[loopangle=-90.0](n4){!grant,$x_{get}\le 2$}

\drawloop(n5){?get}

\drawloop[loopangle=260.0,loopCW=n](n5){}

\drawedge[dash={2.0 2.0 2.0 3.0}{0.0},curvedepth=4.6](n4,n5){?extra }

\drawedge[curvedepth=4.6](n5,n4){!grant,$x_{get}\le 4$,$\varphi_2$}

\drawqbezier[AHnb=0](49.73,-21.92,49.69,-22.13,49.2,-20.85)
\drawqbezier[AHnb=0](57.22,-56.41,57.04,-57.65,58.55,-58.55)
\node[Nframe=n](n6)(43.49,-36.0){Client APECA $Cl$}

\node[Nframe=n](n7)(43.49,-74.0){Access controller APECA $Acc$}

\end{picture}
}\hspace{1cm}\scalebox{0.65}{
\begin{picture}(43,79)(35,-79)

\put(0,-79){}

\node[NLangle=0.0,Nmr=0.0](n6)(20.0,-12.0){00'}
\node[NLangle=0.0,Nmr=0.0](n7)(60.0,-12.0){10'}
\node[NLangle=0.0,Nmr=0.0](n8)(52.0,-28.0){20'}
\node[NLangle=0.0,Nmr=0.0](n9)(20.0,-52.15){01'}
\node[NLangle=0.0,Nmr=0.0](n10)(60.0,-52.15){11'}
\node[NLangle=0.0,Nmr=0.0](n11)(28.0,-32.0){21'}

\drawedge[curvedepth=3.31](n6,n7){get}
\drawedge[curvedepth=3.04](n7,n6){grant,$x_{get}\le 2$}
\drawedge(n7,n8){}
\drawedge[dash={2.0 2.0 2.0 3.0}{0.0}](n7,n10){ extra}
\drawedge[curvedepth=-3.44](n10,n9){}
\drawedge(n10,n11){}
\drawedge[ELpos=66](n10,n8){get, $x_{get}\le 4$, $\varphi_\product$}
\drawedge(n10,n6){}
\drawedge[ELside=r,curvedepth=-3.57](n9,n10){get}
\drawloop[dash={2.0 2.0 2.0 3.0}{0.0},loopangle=13.71](n7){get,$x_{get}\le 1$ }
\drawloop[dash={2.0 2.0 2.0 3.0}{0.0},loopangle=0.0](n10){get,$x_{get}\le 1$ }

\drawqbezier[AHnb=0](56.62,-41.8,53.45,-43.57,52.12,-50.27)
\drawqbezier[AHnb=0](53.97,-12.96,52.74,-16.32,56.89,-18.52)
\drawedge[dash={2.0 2.0 2.0 3.0}{0.0},ELside=r,ELdist=1.35](n6,n9){extra}
\end{picture}}
\vspace{-0.5cm}\\
\begin{minipage}{0.23\textwidth}
\caption{The two APECAs $Cl$ and $Acc$.}\label{fig:example1}
\end{minipage}
\hspace{0.04\textwidth}
\begin{minipage}{0.23\textwidth}
\caption{The parallel composition $Cl\product Acc$.}\label{fig:example2}
\end{minipage}
\vspace{-0.2cm}
\end{wrapfigure}


The main difference to APTAs is that the probability constraints in APECAs are defined on $L$ instead on $2^X_A\times L$, and that we require completeness for the edge function. 
However, completeness is not a restriction, e.g., we model the case where in location $l$ there exists no outgoing $a$-edge by setting 
$T(l,\true,a,\false) = \;?$.
Completing an APECA in this way does not modify its set of implementations.
Note that a similar approach is also used in \cite{Bertrand2012} to obtain completeness for timed modal specifications.
An implementation of an APECA is a \emph{Probabilistic Event-Clock Automaton} $($PECA$)$, which is a probabilistic variant of ECAs. PECAs form a strict subclass of PTAs. Formally, a PECA is a tuple $\mC=(L,A,X_A,AP,V,T,l_0)$, where $L$, $A$, $AP$, $V$ and $l_0$ are as in PTAs,  $X_A$ is as in the APECA, and ${T:L\times CC(X_A)\times A\times \Dist(L)\to \B_2}$ is a two-valued probabilistic edge function.

\section{Refinement}
\label{sec:refinement}

In this section, we define various refinement notions for APTAs and discuss their relationships. More specifically, we define syntactical refinements based on simulation relations and investigate their relationship to semantical refinement (also referred to as \emph{thorough} refinement), i.e., inclusion of sets of implementations. Since our refinement notions for APTAs are based on the refinement notions for APAs~\cite{DKLLPSW11}, we recall relevant definitions now.
%
%
\begin{definition}[Weak APA refinement~\cite{DKLLPSW11}]\label{def:weakref}
Let $\sA_1=(S_1,A,AP,V_1,T_1,s^0_1)$ and $\sA_2=(S_2,A,AP,V_2,T_2,s_2^0)$ be two APAs. A relation $R \subseteq S_1 \times S_2$ is called a \emph{weak refinement relation} iff, for all $(s_1,s_2) \in R$, the following conditions hold: 
\begin{enumerate}
 \item $\forall a\in A, \forall \varphi_2 \in PC(S_2): s_2 \musttrans{a}_2 \varphi_2 \Longrightarrow \exists \varphi_1 \in PC(S_1): s_1 \musttrans{a}_1 \varphi_1$ and $\forall \mu_1 \in \Sat(\varphi_1): \exists \mu_2 \in \Sat(\varphi_2)$ with $\mu_1 \Subset_R \mu_2$ (see the definition of $\Subset_R$ in \cite{HKKG13});
 \item $\forall a \in A, \forall \varphi_1 \in PC(S_1): s_1 \trans{a}_1 \varphi_1 \Longrightarrow \exists \varphi_2 \in PC(S_2): s_2 \trans{a}_2 \varphi_2$ and $\forall \mu_1 \in \Sat(\varphi_1): \exists \mu_2 \in \Sat(\varphi_2)$ with $\mu_1 \Subset_R \mu_2$; and
 \item $V_1(s_1) \subseteq V_2(s_2)$.
 \end{enumerate}
We write $\sA_1 \preceq_W \sA_2$ iff there exists a weak refinement relation relating $s_1^0$~and~$s_2^0$.
\end{definition}
Note that the correspondence function (see \cite{HKKG13}) is not fixed in advance in weak refinements. This is the case in \emph{strong} APA refinements~\cite{DKLLPSW11}, which we denote by $\preceq_S$.

We are now in a position to define our refinement notions for APTAs. While \emph{thorough} refinement is a \emph{semantical} inclusion between sets of implementations, \emph{strong} and \emph{weak} refinements are its \emph{syntactical} counterparts. For the latter two, we apply the refinement notions for APAs to the induced region automata.
%
\begin{definition}[APTA refinements]\label{def:refinement}
Let $\mA_1 = (L_1,A,X,AP,V_1,T_1,l^0_1)$ and $\mA_2 = (L_2,A,X,AP,V_2,T_2,l^0_2)$ be two APTAs. We say that
\begin{enumerate}
  \item $\mA_1$ \emph{thoroughly refines} $\mA_2$, denoted as $\mA_1 \preceq_T \mA_2$, iff $\lb \mA_1 \rb \subseteq \lb \mA_2 \rb$;
\item  $\mA_1$ \emph{\textsf{Sd}-thoroughly refines} $\mA_2$, denoted as $\mA_1 \preceq_T^\textsf{Sd} \mA_2$, iff $\lb \mA_1 \rb^\textsf{Sd} \subseteq \lb \mA_2 \rb^\textsf{Sd}$;
\item  $\mA_1$ \emph{\textsf{Pd}-thoroughly refines} $\mA_2$, denoted as $\mA_1 \preceq_T^\textsf{Pd} \mA_2$, iff $\lb \mA_1 \rb^\textsf{Pd} \subseteq \lb \mA_2 \rb^\textsf{Pd}$;
  \item $\mA_1$ \emph{strongly refines} $\mA_2$, denoted as $\mA_1 \preceq_S \mA_2$, iff $ \sR(\mA_1) \preceq_S \sR(\mA_2)$;
  \item $\mA_1$ \emph{weakly refines} $\mA_2$, denoted as $\mA_1 \preceq_W \mA_2$, iff $ \sR(\mA_1) \preceq_W \sR(\mA_2)$.
  \end{enumerate}
\end{definition}
By Theorem~\ref{thm:apta-apa-reduction}, we can directly obtain that, for any APTAs $\mA_1$ and $\mA_2$, it also holds that: 
\begin{equation}\label{equ:TRefCorr}
\mA_1 \preceq_T \mA_2 \quad \mbox{ \emph{iff} } \quad \sR(\mA_1) \preceq_T \sR(\mA_2)
\end{equation}
where $\sR(\mA_1) \preceq_T \sR(\mA_2)$ refers to thorough refinement for APAs, which is also defined as inclusion of implementation sets~\cite{DKLLPSW11} (analogously for $\preceq_T^\textsf{Sd}$ and $\preceq_T^\textsf{Pd}$). An example of a strong refinement can be found in \cite{HKKG13}.

\noindent
The following theorem establishes a hierarchy among the different notions of refinement. We use $RF_1\,\supset\,RF_2$ to indicate that the refinement $RF_1$ is strictly finer than the refinement $RF_2$. 
\begin{theorem}\label{thm:refHierarchy}
APTA refinements form the following hierarchy:
 $
 \preceq_T^\Sd\ \stackrel{}{\supset}\ \preceq_T^\Pd \ \stackrel{}{\supset}\ \preceq_T  \ \stackrel{}{\supset}\  \preceq_W  \ \stackrel{}{\supset}\ \preceq_S 
 $.
\end{theorem}

\noindent
In Proposition~\ref{prop:PTAB}, we relate weak and strong refinement with probabilistic time-abstracting bisimulation~\cite{CHK08} for PTAs.
In Proposition~\ref{prop:det} we further show that strong, weak and thorough refinements coincide for deterministic APTAs.

\begin{proposition}\label{prop:PTAB}Let $\sim$ denote probabilistic time-abstracting bisimilarity \cite{CHK08} on PTAs. If $\mA_1$ and $\mA_2$ are implementations, then 
\emph{(a)}~$\mA_1\preceq_W \mA_2$ iff $\mA_1 \sim\mA_2$; and
\emph{(b)}~$\mA_1\preceq_S \mA_2$ only if $\mA_1 \sim\mA_2$.
\end{proposition}
%

\begin{proposition}\label{prop:det}
For deterministic APTAs, where the sets of admissible atomic propositions in the initial locations are singletons, thorough, strong and weak refinement coincide.
\end{proposition}

\begin{figure}[h]
\vspace{-0.3cm}
\hspace{1.8cm}\scalebox{0.68}{
\begin{picture}(182,33)(0,-33)\nullfont
\node[Nframe=n](n100)(4.01,-12.12){\begin{large}$\mA_1:$\end{large}}

\node(n0)(39.76,-12.12){$l_1$}

\node(n1)(75.76,-12.12){$l_2$}

\node[Nframe=n](n104)(96.35,-12.12){\begin{large}$\mA_1':$\end{large}}

\node[iangle=90,Nmarks=i](n4)(104.35,-12.12){$l_0$}

\node(n6)(132.11,-12.14){$l_1$}

\drawloop[loopdiam=7,loopangle=90](n6){$b,x<1,\varphi'$ }

\drawedge[dash={3.0 2.0 2.0 2.0}{0.0}](n4,n6){$b,x<1,\varphi'$ }

\node[Nframe=n](n107)(160.85,-12.0){\begin{large}$\mA_2,\mA_2':$\end{large}}

\node[iangle=90,Nmarks=i](n7)(172.85,-12.0){$l_0$}

\node[Nfill=y,fillcolor=Black,Nw=2.0,Nh=2.0,Nmr=1.0](n15)(59.76,-12.12){}

\drawedge[ELside=r,AHnb=0](n0,n15){$a,x<1$}

\drawedge[ELside=r](n15,n1){$1/2$}

\drawedge[ELside=r,ELpos=49,ELdist=0.65,curvedepth=-6.48](n15,n0){$1/2$}

\node[NLangle=0.0,iangle=90,Nmarks=i](n16)(12.01,-12.13){$l_0$}

\drawedge[dash={3.0 2.0 2.0 2.0}{0.0}](n16,n0){$a,x<1,\varphi$}

\end{picture}

}\vspace{-1.2cm}\caption{Counterexamples for showing the \emph{strictly} finer relations}\label{fig:cex}\vspace{-6mm}\end{figure}
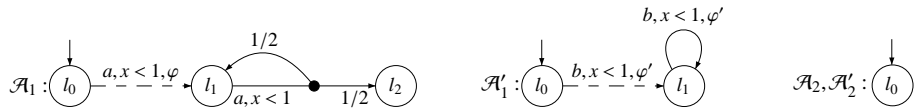

\begin{remark}
 The counterexamples in Fig.~\ref{fig:cex} also show that the \Sd- and \Pd-thorough refinements do not coincide with each other or with thorough refinement even for deterministic APTAs.
\end{remark}

\section{Abstraction}
\label{sec:abst}

The goal of abstraction is to hide internal details of a specification and thereby to obtain
a simpler and usually smaller specification. In the setting of automata, an abstraction 
can be defined by partitioning the state space, i.e., by forming 
disjoint groups of states (or locations) where each of these groups is mapped to one 
abstract state (or location).

Given a set of locations $L$, an \emph{abstraction function} for $L$ is a surjective function \mbox{$\alpha: L \to \tilde{L}$}. 
Its inverse $\gamma: \tilde{L} \to 2^L$ is called a \emph{concretization function}. The abstraction of 
$\mu\in\Dist(2^X\times L)$, denoted $\alpha(\mu)\in\Dist(2^X\times \tilde{L})$ is uniquely defined by 
$\alpha(\mu)(\tilde{l})=\mu(\gamma(\tilde{l}))$, for all $\tilde{l}\in \tilde{L}$. 
Abstraction is lifted to sets of states, sets of distributions, and sets of probability constraints in a 
pointwise manner. It follows that $\tilde{\varphi}=\alpha(\varphi)$ iff $\Sat(\tilde{\varphi})=\alpha(\Sat(\varphi))$. 
The abstraction of the product of constraint function $\varphi$ and $\varphi'$ is given as 
$\alpha(\varphi\cdot\varphi')=\alpha(\varphi)\cdot\alpha(\varphi')$.

A technical challenge in defining abstraction for APTAs is the handling of guards. For this purpose,
we introduce a pre-processing step that syntactically transforms an APTA into an equivalent APTA
such that an abstraction function can be applied. 

\begin{definition}
 Let $\mA = (L,A,X,AP,V,T,l_0)$ be an APTA, $\alpha: L \to \tilde{L}$ be an abstraction function, 
$\gamma: \tilde{L} \to 2^L$ its concretization function. We define the function $\mathrm{g}:\tilde{L}\times A\to CC(X)$ s.t.\  
$\mathrm{g}(\tilde{l},a)= \bigwedge g_i$, if $\forall  l_i\in \gamma(\tilde{l}): \exists \varphi_i \in PC(2^X\times L), g_i\in CC(X): T(l_i,g_i,a,\varphi_i) = \top$; and $ \mathrm{g}(\tilde{l},a)=\false$, otherwise.
\end{definition}
\noindent Here, $\mathrm{g}$ calculates the common guards of all must-transitions emitting from $\tilde{l}$ with action $a$. 
Given the function $\mathrm{g}$, we define a pre-processing step for APTAs that splits some of the must-transitions such that 
abstraction has the intended meaning. 

%
%
Given a guard $g \in CC(X)$, we define the negation of $g$, denoted $\bar{g} \subseteq CC(X)$, as the set of guards, such that for any valuation $v\in \R^X$ it holds that 
$v \triangleright g$ if and only if there exists no $g' \in \bar{g}$ such that $v \triangleright g'$. The negation of a guard can be effectively computed by splitting the guard into its atomic comparisons and inverting the
comparisons.

\begin{definition}[Pre-processing]\label{def:preprocess}
Let $\mA = (L,A,X,AP,V,T,l_0)$ be an APTA, $\alpha: L \to \tilde{L}$ be an abstraction function, 
$\gamma: \tilde{L} \to 2^L$ be its concretization function. Let $\mathrm{g}:\tilde{L}\times A\to CC(X)$ 
be defined as before. The pre-processing function $\mathcal{P}_\alpha$ maps $\mA$ to the APTA $\mP_\alpha(\mA)=(L,A,X,AP,V,T',l_0)$ such that for any 
$l\in L$, $g\in CC(X)$ and $\varphi\in PC(2^X\times L)$, if $T(l,g,a,\varphi)=\top$, then 
$T'(l,\mathrm{g}(\alpha(l),a),a,\varphi)=\top$ and $\forall g' \in \bar{\mathrm{g}}(\alpha(l),a)$:
$T'(l,g \wedge g',a,\varphi_i)=\top$; 
and $T'(l,g,a,\varphi)=T(l,g,a,\varphi)$, otherwise. 
\end{definition}
\noindent As a result of the pre-processing function, the guards on a must-transition are either the common guard determined by $\mathrm{g}$, or are disjoint with the common guard. Since $\mathrm{g}(\alpha(l),a)$ and $g\wedge g'$ for all $g'\in\bar{\mathrm{g}}(\alpha(l_i),a)$ form a partition of $g$, it is easy to see that $\sR(\mA)=\sR(\mathcal{P}_\alpha(\mA))$. 
We are now in a position to define the abstraction.

\begin{definition}[APTA Abstraction]
\label{def:abstr3}
Given an abstraction function  $\alpha: L \to \tilde{L}$ and 
its concretization function $\gamma: \tilde{L} \to 2^L$, a pre-processed APTA $\mP_\alpha(\mA) = (L,A,X,AP,V,T,l_0)$ and a guard function $\mathrm{g}:\tilde{L}\times A\to CC(X)$. Let 
$\alpha(\mP_\alpha(\mA)) = (\tilde{L},A,X,AP,\tilde{V},\tilde{T},\alpha(l_0))$ be the APTA defined by: $\tilde{V}(\tilde{l}) = \bigcup_{l\in \gamma(\tilde{l})} V(l)$ and 

\noindent$\tilde{T}(\tilde{l},\tilde{g},a,\tilde{\varphi})=\left\{ 
  \begin{array}{cl}
  \top & \;\; \text{if } \tilde{g}=\mathrm{g}(\tilde{l},a), \text{and }  \Sat(\tilde{\varphi}) = \alpha(\bigcup_{\la l,\varphi \ra \in \gamma(\tilde{l})\times PC(2^X\times L): T(l,\tilde{g},a,\varphi) = \top} \Sat(\varphi))\\
   ?   & \;\; \text{if } \tilde{g} \neq\mathrm{g}(\tilde{l},a), \text{and }\exists l\in \gamma(\tilde{l}),\varphi \in PC(2^X\times L) : T(l,\tilde{g},a,\varphi) \neq \bot, \text{and} \\
       & \quad \Sat(\tilde{\varphi}) = \alpha(\bigcup_{\la l,\varphi \ra \in \gamma(\tilde{l}) \times PC(2^X\times L): T(l,\tilde{g},a,\varphi)  \neq \bot} \Sat(\varphi))\\
%
  \bot & \;\; \text{otherwise}
  \end{array}
  \right.$
\end{definition}

\begin{lemma}
\label{lem:abstraction}
Let ${\alpha}(\mP_\alpha(\mA))$ be an abstraction of $\mA$. Then there exists an APA abstraction function (cf.\,\cite{DKLLPSW11})
$\alpha'$ on $\sR(\mA)$, such that $\sR({\alpha}(\mP_{\alpha}(\mA)))={\alpha'}(\mP_{\alpha'}(\sR(\mA)))$. 
\end{lemma}
\begin{proposition}\label{prop:absRef}
For any APTA $\mA$ and abstraction function $\alpha$, $\mA \preceq_W {\alpha}(\mP_\alpha(\mA))$.
\end{proposition}
Lemma~\ref{lem:abstraction} states that an abstraction function for an APTA $\mA$ induces an abstraction function on $\sR(\mA)$. Proposition~\ref{prop:absRef} follows directly from Lemma~\ref{lem:abstraction} and a similar result 
known for APAs~\cite{DKLLPSW11}. This result is important in order to ensure that applying an abstraction yields 
a generalized specification, i.e., formally that the original specification always weakly refines its abstraction.
Note also that we show in Section~\ref{sec:comp} that abstraction interacts well with parallel composition. 

\begin{figure}[!b]\subfigure[APECA $Cl_1$.]{
\qquad\quad\scalebox{0.65}{
\begin{picture}(95,48)(0,-48)
 \put(0,-48){}
\node[NLangle=0.0](n0)(8.0,-20.0){0'}

\node[NLangle=0.0](n1)(43.48,-20.0){1'}

\node[NLangle=0.0](n2)(83.23,-20.0){1''}

\node[NLangle=0.0](n3)(23.74,-40.0){2'}

\drawedge(n0,n1){!get, $x_{get}<2$}

\drawedge[curvedepth=3.88](n1,n2){!get, $x_{get}\ge 2$}

\drawedge[curvedepth=3.61](n2,n1){}

\drawedge[curvedepth=7.24](n2,n3){}

\drawedge[curvedepth=10.56](n2,n0){?grant, $\varphi_3$}

\drawloop(n1){!extra}

\drawloop[dash={3.0 2.0 2.0 2.0}{0.0}](n2){!extra}

\drawloop[dash={3.0 2.0 2.0 2.0}{0.0},ELside=r,loopangle=-90.0,loopCW=n](n2){!get $x_{get}\le 2$}

\drawqbezier[AHnb=0](76.73,-21.39,76.38,-22.64,79.14,-23.8)
\drawedge[curvedepth=16.98](n0,n2){!get, $x_{get}\ge 2$}

\drawloop(n0){!extra,$x_{extra}\ge 1$}

\end{picture}
}}\qquad\quad\subfigure[The abstraction $\alpha'(Cl_1)$.]{\scalebox{0.65}{
\begin{picture}(95,48)(0,-48)
 \put(0,-48){}
\node(n1)(24.0,-20.0){0}

\node(n2)(64.0,-20.0){1''}

\node(n3)(24.0,-44.0){2'}

\drawedge[curvedepth=4.2](n1,n2){!get, $x_{get}\ge 2$}

\drawedge[ELpos=60,ELdist=2.61,curvedepth=4.2](n2,n1){?grant,$\varphi_3'$}

\drawedge(n2,n3){}

\drawloop[dash={2.0 2.0 2.0 3.0}{0.0}](n1){!extra, $x_{extra}<1$}

\drawloop[dash={2.0 2.0 2.0 3.0}{0.0},loopangle=-90.0](n1){!get, $x_{get}< 2$}

\drawloop[dash={2.0 2.0 2.0 3.0}{0.0},loopangle=-90.0](n2){!get, $x_{get}\le 2$}

\drawloop[dash={2.0 2.0 2.0 3.0}{0.0}](n2){!extra  }

\drawqbezier[AHnb=0](51.87,-27.54,50.44,-27.3,49.2,-23.53)
\drawloop[loopangle=180](n1){\begin{minipage}{1.6cm}
                              !extra,\\ $x_{extra}\ge 1$
                             \end{minipage}}
\end{picture}}
}\vspace{-3mm}\caption{APECA abstraction.}\label{fig:abstraction}\vspace{-3mm}
\end{figure}
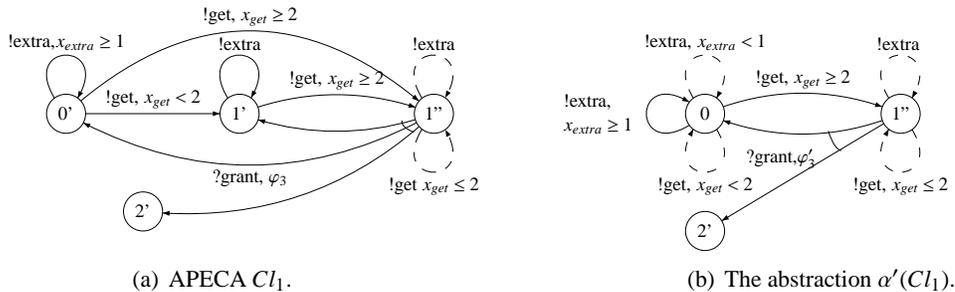

\begin{example}\label{ex:abstraction}
Another client specification $Cl_1$ is depicted in Fig.\,\ref{fig:abstraction}(a), where state~1 in 
$Cl$ is split into 1' and 1'' in $Cl_1$. State 1' can be seen as a ``quick phase'', since a \emph{get} 
request should be sent in less than $2$ time units, while state 1'' is the ``slow phase'' due to the guard 
$x_{get}\ge 2$. Furthermore, in state 0' and 1', an extended time slot will be granted whether needed or not. 
However, in state 0', the additional time will only be granted after 1 time unit. From state 1'', 
it is possible to either start from 0' again (quick phase), to move to state 1' (slow phase), or 
to state 2' (termination) after the access to the resource has been granted.


Let the abstraction function be defined as $\alpha(l)=0$ for $l\in \{0',1'\}$ and $\alpha(l)=l$ for $l\in\{1'',2'\}$. 
The pre-processing splits the edge (!\emph{extra},\emph{true}) from state 1' into (!\emph{extra}, $x_{extra}\ge 1$) and (!\emph{extra}, $x_{extra}< 1$). 
The abstraction ${\alpha}(Cl_1)$ is shown in Fig.\,\ref{fig:abstraction}(b). 
Any distribution $\mu'$ satisfying the constraint $\varphi_3'$ in $\alpha'(Cl_1)$ 
is defined such that $\mu'(0)=\mu(0')+\mu(1')$, for any $\mu\in \Sat(\varphi_3)$ in $Cl_1$. 
\end{example}


\section{Conjunction and Parallel Composition}
\label{sec:comp}

In this section, we define two composition operators for APECAs, i.e., conjunction and parallel composition. These two operators are intentionally defined only for APECAs, and not for general APTAs, in order to be able to ensure compositionality properties. Conjunction and parallel composition form the cornerstones for a specification theory supporting independent development and structural composition.

\subsection{Conjunction}
Given two APECAs $\mE_1$ and $\mE_1$, their conjunction (or \emph{logical composition}), denoted as
$\mE_1\wedge \mE_2$, is the specification that realizes the conjunctive behavior of $\mE_1$ and $\mE_2$.
Specifically, the set of implementations of the conjunction approximates the intersection of the sets of 
implementations of $\mE_1$ and $\mE_2$. We define conjunction here only for action-deterministic APECAs 
over the same set of actions. For two automata with different action sets, we add the missing actions
in the respective automaton, and complete the edge function as explained in Section~\ref{sec:apeca}. 
We leave the generalization to nondeterministic APECAs for future work.

\begin{definition}[APECA Conjunction]\label{def:conjunction2}
Let $\mE_1=(L_1,A,X_A,AP,V_1,T_1,l_0^1)$ and $\mE_2=(L_2,A,X_A,AP,$ $V_2,T_2,l_0^2)$ be two action-deterministic 
APECAs over the same sets of actions and atomic propositions. 
Their \emph{conjunction} is 
\halflineup
\[ 
\mE_1 \wedge \mE_2= (L_1 \times L_2, A, X_A, AP, V_\wedge, T_\wedge, \la l_0^1,l_0^2\ra)
\] 
where $V_\wedge(\la l_1,l_2 \ra) = V_1(l_1) \cap V_2(l_2)$ for all $l_1 \in L_1,l_2\in L_2$ and 
$T_\wedge$ is defined as follows: 
for all $a \in A$, $g_1,g_2\in CC(X_A)$, and $l_1 \in L_1$, $l_2 \in L_2$:
   \begin{enumerate}
   \item \label{itm:conj}
   $\forall \varphi_1 \in PC(L_1)$,  
   $\forall \varphi_2 \in PC(L_2)$ such that 
   $T_1(l_1,g_1,a,\varphi_1) \neq \bot$ and $T_2(l_2,g_2,a,\varphi_2) \neq \bot$, let 
   $ T_\wedge(\la l_1,l_2 \ra,$ $g_1\wedge g_2,a,\varphi_\wedge) = 
    T_1(l_1,g_1,a,\varphi_1) \;\sqcup\; T_2(l_2,g_2,a,\varphi_2) $ with $\varphi_\wedge$ the new constraint in 
    $PC(L_1 \times L_2)$ such that $\mu_\wedge \in \Sat(\varphi_\wedge)$ iff
   \begin{itemize}
     \item the distribution {$\mu_1  = \{\; k_1 \mapsto \sum_{k_2 \in L_2} \mu_\wedge(\la k_1,k_2 \ra) \;\} $} is in $\Sat(\varphi_1)$, and
     \item the distribution {$\mu_2= \{\; k_2 \mapsto \sum_{k_1 \in L_1}  \mu_\wedge(\la k_1,k_2 \ra) \;\} $} is in $\Sat(\varphi_2)$.
   \end{itemize}
   \item For all other $\varphi'_\wedge \in PC(L_1 \times L_2)$ and $g' \in CC(X_A)$ we define 
   $T_\wedge(\la l_1,l_2 \ra, g',a,\varphi'_\wedge) = \bot$.
   \end{enumerate}
 \end{definition}

\noindent
Informally, the conjunction $\mE_1 \wedge \mE_2$ can be regarded as the \emph{largest} specification that refines $\mE_1$ \emph{and} $\mE_2$.
Note that we rely on the completeness of the edge functions. For example, assume that in $l_1$ there is a 
must-edge via the action $a$, and in $l_2$ the action $a$ is not enabled. 
As we require completeness, the latter means that there is an edge $l_2 \maytrans{g,a} \false$. In the conjunction, this edge
is combined with the must-edge from $l_1$ using Rule~\ref{itm:conj}. Thus, in the conjunction we obtain the edge 
$\la l_1,l_2 \ra \musttrans{g,a} \false$, which means that $\la l_1,l_2 \ra$ is inconsistent. 
On the other hand, for may-edges no inconsistencies are produced. 

\begin{figure}[!ht]
\vspace{-4mm}
\subfigure[APECA $Cl_2$.]{\hspace{-0.4cm}
\qquad\qquad\scalebox{0.65}{
\begin{picture}(95,48)(0,-48)
\put(0,-48){}
\node[NLangle=0.0](n0)(8.0,-20.0){0'}

\node[NLangle=0.0](n1)(36.0,-20.0){1'}

\node[NLangle=0.0](n2)(68.0,-20.0){1''}

\node[NLangle=0.0](n3)(20.0,-44.0){2'}

\drawedge[curvedepth=3.88](n0,n1){!get}

\drawedge[curvedepth=3.88](n1,n2){!get}

\drawedge[curvedepth=3.61](n2,n1){}

\drawedge[curvedepth=7.24](n2,n3){}

\drawedge[curvedepth=10.56](n2,n0){?grant, $\varphi_4$}

\drawloop(n1){!extra}

\drawloop[dash={2.0 2.0 2.0 3.0}{0.0}](n2){!extra}

\drawloop[dash={2.0 2.0 2.0 3.0}{0.0},ELside=r,loopangle=-90,loopCW=n](n2){!get $x_{get}\le 2$}

\drawqbezier[AHnb=0](60.96,-22.46,60.43,-24.42,63.37,-25.13)

\end{picture}
}}\qquad \subfigure[APECA $Cl\wedge Cl_2$.]{\hspace{-1.3cm}\scalebox{0.65}{
\begin{picture}(95,48)(0,-48)
\put(0,-48){}
\node[NLangle=0.0](n12)(12.0,-20.0){00'}

\node[NLangle=0.0](n13)(52.0,-20.0){11'}

\node[NLangle=0.0](n14)(92.52,-20.0){11''}

\node[NLangle=0.0](n15)(12.0,-44.0){02'}

\node[NLangle=0.0](n16)(32.0,-44.0){20'}

\node[NLangle=0.0](n17)(52.0,-44.0){21'}

\node[NLangle=0.0](n18)(72.0,-44.0){22'}

\node[NLangle=0.0](n19)(92.52,-44.0){01'}

\drawedge[curvedepth=4.01](n12,n13){!get}

\drawedge[curvedepth=4.01](n13,n14){!get,$x_{get}\le 1$}

\drawedge[curvedepth=-2.54](n14,n19){}

\drawedge[curvedepth=-2.81](n19,n14){}

\drawedge(n14,n18){}

\drawedge(n14,n17){}

\drawedge(n14,n16){}

\drawedge(n14,n15){?grant,$\varphi_\wedge$}

\drawedge[curvedepth=5.08](n14,n12){}

\drawqbezier[AHnb=0](86.09,-20.85,85.2,-24.6,90.37,-26.74)
\drawloop(n13){!extra}

\drawloop[loopangle=-2.94](n19){!extra}

\drawloop[dash={2.0 2.0 2.0 3.0}{0.0}](n14){!get,$x_{get}\le 1$}

\drawloop[dash={2.0 2.0 2.0 3.0}{0.0},loopangle=-3.01](n14){!extra}

\end{picture}
}}
\vspace{-1mm}
\caption{APECA conjunction.}\label{fig:conjunction}
\vspace{-3mm}
\end{figure}
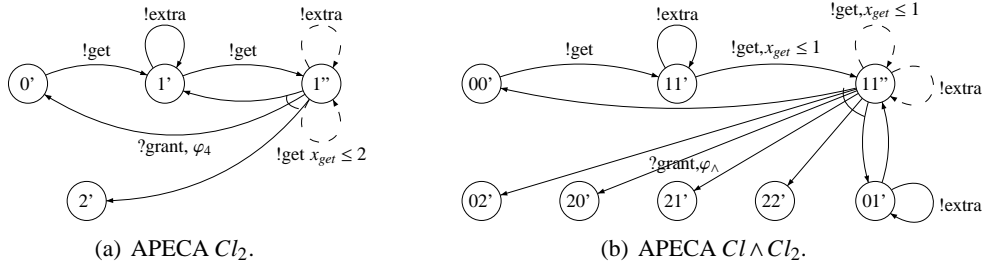

\begin{example}\label{ex:conjunction}
 Fig.\,\ref{fig:conjunction}(a) depicts $Cl_2$ as another version of the client. 
 The conjunction of $Cl$ and $Cl_2$ is shown in Fig.\,\ref{fig:conjunction}(b). Let $\varphi_1$ be defined as in Example\,\ref{ex:apeca} and let the probability from state 1'' to state 0' be $q_1$ and from state 1'' to state 1' be $q_2$ and from state 1'' to 2' be $q_3$ in $Cl_2$. 
We set $\varphi_4$ such that $0\le q_1\le 1/5$, $1/3\le q_2\le 1$ and $q_1+q_2+q_3=1$. Now let $\mu_\wedge$ be any distribution satisfying $\varphi_\wedge$. Let $\mu_\wedge(00')=r_1$, $\mu_\wedge(01')=r_2$, $\mu_\wedge(02')=r_3$, $\mu_\wedge(20')=r_4$, $\mu_\wedge(21')=r_5$, $\mu_\wedge(22')=r_6$. Those $r$'s should satisfy the following constraints, given the constraints on the $p$'s and the $q$'s:
\vspace{-0.2cm}
\begin{align*}
 r_1+r_2+r_3=p_1\quad r_4+r_5+r_6=p_2\quad r_1+r_4=q_1\quad r_2+r_5=q_2\quad r_3+r_6=q_3.
\end{align*}
\end{example}

We use the notation $\sA_1 \equiv \sA_2$ to denote that
both $\sA_1 \preceq_W \sA_2$ and \mbox{$\sA_2 \preceq_W \sA_1$}.
Next, we show that conjunction is preserved by the region construction.

\begin{lemma}\label{lem:wedgeR}
 For any APECAs $\mE_1, \mE_2$, it holds that $\sR(\mE_1 \wedge\mE_2)\equiv\sR(\mE_1)\wedge\sR(\mE_2)$.
\end{lemma}




\noindent
We now show that conjunction is the greatest lower bound of APECAs w.r.t.\ weak refinement,
after pruning the conjunction. Note that $\beta^*(\mE)$ means applying the pruning operator $\beta$ on $\mE$ for finitely many times. Theorem~\ref{thm:conjRef} relates the sets of implementations of the 
conjunction with the implementation sets of its \mbox{components}.

\begin{theorem}\label{thm:conjunction1}
Let $\mE_1$, $\mE_2$ and $\mE_3$ be action-deterministic consistent APECAs. Then the following properties hold:
$
  (1)~\beta^*(\mE_1 \wedge \mE_2) \preceq \mE_1;
  \quad (2)~\text{if } \mE_3 \preceq \mE_1 \text{ and } \mE_3 \preceq \mE_2\text{, then }\mE_3 \preceq \beta^{*}(\mE_1 \wedge \mE_2). 
$
\end{theorem}

%

\begin{theorem}\label{thm:conjRef}
For any APECAs $\mE_1$, $\mE_2$ it holds: $\lb \mE_1 \wedge \mE_2 \rb \subseteq \lb \mE_1 \rb \cap \lb \mE_2 \rb$.
\end{theorem}

\subsection{Parallel Composition}

We now define a parallel composition operator for APECAs which enables modular specifications of systems.
The formal definition is similar to the one for conjunction and requires again that all 
actions are shared. To enable interleaving of non-shared actions, we realize completeness
of the edges differently. Let $a$ be an action of $\mE_1$ that does not occur
in $\mE_2$. We now add $a$ to the set of actions of $\mE_2$ and for every location $l$ in $\mE_2$ we add a
self-loop must-edge such that $T_2(l,\true,a,\varphi)$ with $\Sat(\varphi) = \{ \mu_l \}$, where
$\mu_l$ is the point distribution on $l$. This is repeated for all non-shared actions of $\mE_2$ 
(symmetrically for $\mE_1$) until $\mE_1$ and $\mE_2$ have the same set of actions.
This simplifies the definition of the parallel composition (and our proofs for related theorems)
since interleaved and synchronized behavior can be uniformly treated using a single composition~rule.   

\begin{definition}[APECA Parallel Composition] \label{def:par}
Given two APECAs 
$\mE_1 = (L_1,A,X_A,AP_1,V_1,T_1,l^1_0)$,
$\mE_2 = (L_2,A,X_A,AP_2,V_2,T_2,l^2_0)$ with
$AP_1 \cap AP_2 = \emptyset$. The \emph{parallel composition} of $\mE_1$ 
and $\mE_2$, written as 
$\mE_1 \product \mE_2$, is defined as 
\[
\mE_1 \product \mE_2 = (L_1\times L_2,A,X_A, AP_1 \cup AP_2, V_\product,T_\product,\la l^1_0,l^2_0\ra) 
\quad\text{ where}
\]
\begin{itemize}
  \item $T_\product$ is defined as follows. For all $l_1 \in L_1$, $l_2 \in L_2$, $g_1,g_2 \in CC(X_A)$, $a \in A$:
  \begin{itemize}
  \item 
  $\forall \varphi_1 \in PC(L_1)$ and 
  $\forall \varphi_2 \in PC(L_2)$ such that 
 $T_1(l_1,g_1,a,\varphi_1) \neq \bot$ and $T_2(l_2,g_2,a,\varphi_2) \neq \bot$, let
  $
  T_\product(\la l_1,l_2 \ra, g_1\wedge g_2,a,\varphi_\product) = T_1(l_1,g_1,a,\varphi_1) \sqcap T_2(l_2,g_2,a,\varphi_2)
  $
  with $\varphi_\product$ the new constraint on $\mA_1 \product \mA_2$ defined as follows:  
  $\mu_\product \in \Sat(\varphi_\product)$ if and only if there exist 
  $\mu_1 \in \Sat(\varphi_1)$ and 
  $\mu_2 \in \Sat(\varphi_2)$ s.t.\ 
  $
  \mu_\product(\la k_1,k_2 \ra) = \mu_1(k_1) \cdot \mu_2(k_2) \;\text{ for all } k_1\in L_1, k_2\in L_2.
  $
\item
  For all other $\varphi'_\product \in PC(L_1 \times L_2)$ and $g' \in CC(X_A)$ we define 
  $T_\product(\la l_1,l_2 \ra, g',a,\varphi'_\product) = \bot$.
  \end{itemize}
  \item $V_\product(\la l_1,l_2\ra) = \{ E_1 \cup E_2 \;|\; E_1\in V_1(l_1) \wedge E_2\in V_2(l_2)\}$ for all $l_1 \in L_1, l_2 \in L_2$.
  \end{itemize}
\end{definition}


\begin{example}
\label{ex:parallel}
Fig.\,\ref{fig:example2} shows the parallel composition $Cl \product Acc$ of the client and access controller in Fig.\,\ref{fig:example1}.
%
Let $\varphi_1$ be defined as in Example \ref{ex:conjunction}. Let the probability from state 1' to state 0' be $p_3$ and from state 1' to state 1' be $p_4$ in $Acc$. 
Set $\varphi_2$ be $0\le p_3\le 1/2$ and $p_3+p_4=1$. In  
$Cl\product Acc$, let the probability from state 11' to states 00', 01', 20' and 21' be 
$q_1$, $q_2$, $q_3$ and $q_4$, respectively. The resulting $\varphi_\product$ is $q_1+q_2+q_3+q_4=1$ 
and $q_1=p_1\cdot p_3$, $q_2=p_1\cdot p_4$, $q_3=p_2\cdot p_3$ and $q_4=p_2\cdot p_4$. 
\end{example}

\noindent
Similarly to conjunction, parallel composition is preserved by the region construction 
(Lemma~\ref{lem:paraR}). Moreover, parallel composition interacts well with refinement
(Theorem~\ref{thm:parRef}) and abstraction (Theorem~\ref{thm:absPar}). The latter result
allows us to avoid state space explosion by applying abstraction component-wise instead
of computing the complete system specification and then applying the abstraction.
By Theorem~\ref{thm:parRef} we also know that $\equiv$ is a congruence w.r.t.\ parallel composition.

\begin{lemma}\label{lem:paraR}
For any APECAs $\mE_1, \mE_2$ it holds: $\sR(\mE_1 \product \mE_2)\equiv\sR(\mE_1)\product \sR(\mE_2)$.
\end{lemma}


\begin{theorem}\label{thm:parRef}
Weak refinement is a precongruence w.r.t.\ parallel composition.
\end{theorem}

\begin{theorem}\label{thm:absPar}
Let $\mE_1$ and $\mE_2$ be APECAs and $\alpha_1,\alpha_2$ be abstraction functions. 
The following equalities hold up to isomorphism: 
\begin{enumerate}\item $\mP_{\alpha_1}(\mE_1)\product \mP_{\alpha_2}(\mE_2) = \mP_{\alpha_1\times\alpha_2}(\mE_1\product\mE_2)$; 
 \item $\alpha_1(\mP_{\alpha_1}(\mE_1)) \product \alpha_2(\mP_{\alpha_2}(\mE_2)) = ({\alpha}_1 \times {\alpha}_2)(\mP_{\alpha_1}(\mE_1) \product  \alpha_2(\mP_{\alpha_2}(\mE_2)) )$;
\item $(\alpha_1\circ\mP_{\alpha_1})(\mE_1) \product (\alpha_2\circ\mP_{\alpha_2})(\mE_2) = ({\alpha}_1 \times {\alpha}_2)\circ(\mP_{\alpha_1\times\alpha_2})(\mE_1 \product \mE_2)$.
\end{enumerate}
\end{theorem}

\vspace{-4mm}
\section{Conclusions}\label{sec:conclusions}

We introduced a modal specification theory of abstract probabilistic timed automata (APTAs) that capture systems which contain nondeterminism, probability and time.
The theory supports refinement, abstraction, and the operations of parallel composition and conjunction, the latter for independent development.
The main challenge in combining probabilistic and timed behaviour was due to the need to consider guards of transitions, as well as a time-divergence sensitive consistency check.
In order to obtain a compositional theory, we had to restrict to the class of abstract probabilistic event clock automata (APECAs).
%
%
%
As future work, we aim at finding a class of modal specifications for probabilistic timed systems
that, in terms of expressiveness, lies between APECAs and APTAs, but still enjoys the compositionality
properties of APECAs. 
Moreover, we plan to extend our work on abstraction to support counterexample generation.

\nocite{CHK08}

\vspace{-0.1cm}
\paragraph*{Acknowledgement. }
This work is supported by the ERC Advanced Grant VERIWARE and EU project CONNECT.
\bibliographystyle{eptcs}
\bibliography{biblio1,biblio}

\end{document}